\documentclass{article}
\usepackage{spconf,amsmath,graphicx}
\usepackage{url, times, booktabs, tabularx, amssymb, bm, mathrsfs, mathtools, bbm}
\usepackage{calrsfs, color}
\usepackage[utf8]{inputenc}
\usepackage[final]{microtype}
\usepackage{booktabs}
\usepackage{array}
\usepackage{enumitem}
\usepackage{soul}
\usepackage{cite}
\usepackage{siunitx}
\usepackage{adjustbox}
\usepackage{hhline}
\usepackage{makecell}
\usepackage{caption}
\usepackage{float}
\usepackage{tablefootnote}
\usepackage{scrextend}
\usepackage{multirow}

\title{A Track-Wise Ensemble Event Independent Network \\for Polyphonic Sound Event Localization and Detection}
\name{Jinbo Hu$^{1,2}$, Yin Cao$^{3}$, Ming Wu$^{1}$, Qiuqiang Kong$^{4}$, Feiran Yang$^{1}$,  Mark D. Plumbley$^{3}$, Jun Yang$^{1,2}$}
\address{ $^{1}$Key Laboratory of Noise and Vibration Research, Institute of Acoustics, \\
Chinese Academy of Sciences, Beijing, China \{hujinbo, mingwu, feiran, jyang\}@mail.ioa.ac.cn\\
$^{2}$University of Chinese Academy of Sciences, Beijing, China\\
$^{3}$Centre for Vision, Speech and Signal Processing (CVSSP), University of Surrey, UK \\
\{yin.cao, m.plumbley\}@surrey.ac.uk \\
$^{4}$ByteDance Shanghai, China, kongqiuqiang@bytedance.com\\
}
\begin{document}
%
\maketitle
\begin{abstract}

Polyphonic sound event localization and detection (SELD) aims at detecting types of sound events with corresponding temporal activities and spatial locations. In this paper, a track-wise ensemble event independent network with a novel data augmentation method is proposed. The proposed model is based on our previous proposed Event-Independent Network V2 and is extended by conformer blocks and dense blocks. The track-wise ensemble model with track-wise output format is proposed to solve an ensemble model problem for track-wise output format that track permutation may occur among different models. The data augmentation approach contains several data augmentation chains, which are composed of random combinations of several data augmentation operations. The method also utilizes log-mel spectrograms, intensity vectors, and Spatial Cues-Augmented Log-Spectrogram (SALSA) for different models. We evaluate our proposed method in the Task of the L3DAS22 challenge and obtain the top ranking solution with a location-dependent F-score to be 0.699. Source code is released\footnote{\url{https://github.com/Jinbo-Hu/L3DAS22-TASK2}}.


\end{abstract}
\section{Introduction}
\label{sec:intro}

Sound event localization and detection (SELD) contains two subtasks, sound event detection (SED) and direction-of-arrival (DoA) estimation. SED aims at detecting types of sound and their corresponding temporal activities. Whereas DoA estimation predicts spatial trajectories of different sound sources. SELD characterizes sound sources in a spatial-temporal manner that can be used in a wide range of applications, such as robot auditory, surveillance, and smart home.

SELD has received broad attention recently. Adavanne et al. proposed a polyphonic SELD work using an end-to-end network, SELDnet, which was utilized for a joint task of SED and regression-based DoA estimation\cite{Adavanne2018_JSTSP}. SELD was then introduced in the Task 3 of the Detection and Classification of Acoustics Scenes and Events (DCASE) 2019 Challenge for the first time, which uses the TAU Spatial Sound Events 2019 dataset\cite{Adavanne2018_JSTSP,Adavanne2019_DCASE}. The Learning 3D Audio Sources (L3DAS) project
held the L3DAS21 and L3DAS22 challenges in 2021. The main novelty is to use two Ambisonics microphone arrays.

SELDnet employed multi-channel magnitude and phase spectrograms as input features\cite{Adavanne2018_JSTSP}. Subsequently, multi-channel log-mel spectrograms, intensity vectors (IV) in log-mel space for first-order Ambisonics (FOA) format signals, and generalized cross-correlation with phase transform (GCC-PHAT) for microphone array (MIC) signals were demonstrated to be more effective features for SELD\cite{Cao2019,cao2020event,wang2021four,perotin2019crnn, celsi2020quaternion}. Nguyen et al. proposed a novel feature called Spatial Cue-Augmented Log-Spectrogram (SALSA), which was composed of multi-channel log spectrograms stacked with normalized principal eigenvectors of a spatial covariance matrix.\cite{nguyen2021dcase,nguyen2021salsa}.

SELDnet has the limitation that it is unable to detect sound events of the same type but with different locations\cite{Adavanne2018_JSTSP}. Event independent network (EIN) with track-wise output format was proposed to tackle this problem\cite{cao2020event,cao2021}. In EIN, there are several event-independent tracks, which means the prediction on each track can be of any event type. The number of tracks needs to be pre-determined according to the maximum number of overlapping events. 

EINV2 utilizes multi-head self-attention (MHSA) to achieve better performance compared with SELDnet\cite{cao2021}. Other network structures, such as DenseNet\cite{huang2017densely} and Conformer\cite{gulati2020conformer}, can also be employed. Our proposed model uses DenseNet and Conformer to extend EINV2.

To further improve the performance of trained models, post-processing methods can be utilized during inference. A spatial augmentation technique is utilized for test-time augmentation (TTA)\cite{nguyen2021dcase,nguyen2021salsa}. Test samples are augmented by 16-pattern rotations\cite{mazzon2019first}. Output is then computed by a mean of all 16 outputs. Average ensemble and weighted ensemble compute a mean or weighted mean output of several different trained models\cite{nguyen2021dcase,shimada2021ensemble}. Stacking is also an ensemble method. It is to train inhomogeneous models using the original dataset at first, and then train an ensemble model using predictions of these inhomogeneous models\cite{Cao2019,wolpert1992stacked}.

In this paper, we propose an ensemble Event Independent Network based on previously proposed EINV2 and a novel data augmentation approach. The method utilizes log-mel, IV, and SALSA featrues for different models. Our proposed model exploits EINV2, combining a track-wise-output format, permutation-invariant trainint (PIT), and soft parameter-sharing (PS). The Conformer structure is employed to learn local and global patterns. The DenseNet structure is utilized to increase the diversity of different models for ensemble. The proposed data augmentation method is characterized by utilizing several augmentation operations. These data augmentation operations are sampled, layered, and combined randomly to produce a high diversity of augmented features. We propose a track-wise ensemble method for the track-wise output format to improve the system performances. The proposed system obtained a Top ranking in the Task 2 of the L3DAS22 challenge\cite{l3das22}.




\section{The Method}
\label{sec:pagestyle}

\subsection{Input Features}
The dataset provided by the L3DAS22 Task 2 uses two FOA microphone arrays that is placed in the center of a room.
In this paper, two types of features are used for ensemble models. Log-mel spectrograms and IV in log-mel space from first-order Ambisonics (FOA) are used as the first set of features. SALSA is used as the second set of features. We extract features for both FOA and concatenate them as input features.

Log-mel spectrograms are first used for SED, while IV in log-mel space is used for DoA estimation\cite{Cao2019,cao2020event,wang2021four}. FOA includes four channels of signals, i.e., omni-directional channel $\mathbf{w}$, and three directional channels $\mathbf{x}$, $\mathbf{y}$, and $\mathbf{z}$. Log-mel spectrograms are computed from the short-time Fourier transform spectrograms of four-channel signals, and intensity vectors are computed from cross-correlation of $\mathbf{w}$ with $\mathbf{x}$, $\mathbf{y}$ and $\mathbf{z}$ in log-mel space.




SALSA is composed of two major components: multi-channel log-linear spectrograms and normalized principal eigenvectors. The detailed information can be found in\cite{nguyen2021salsa,nguyen2021dcase}.

\subsection{Network Architecture}

\begin{figure}[tb]
  \centering
  \scalebox{1.0}{\centerline{\includegraphics[width=\columnwidth]{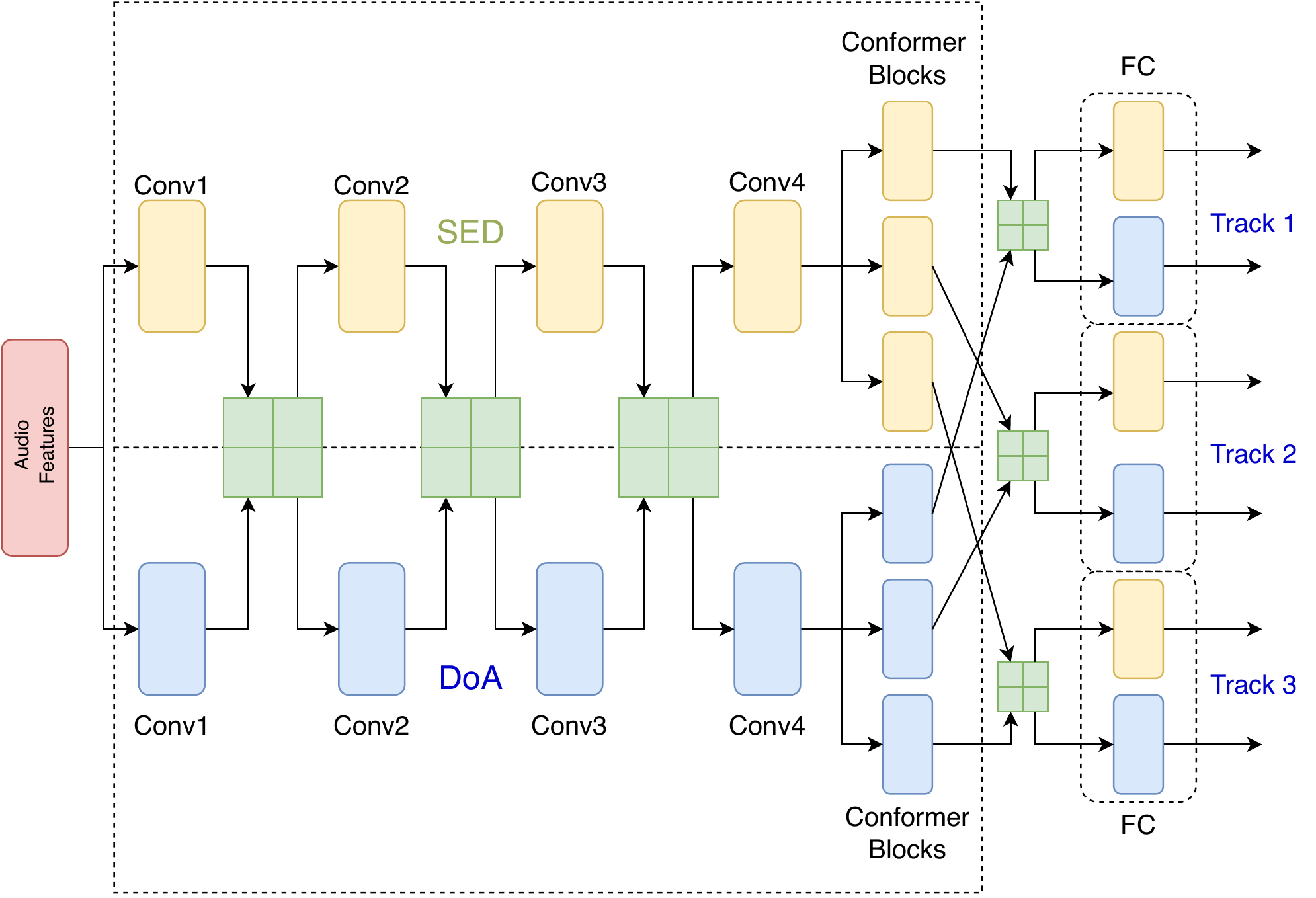}}}
  \vspace*{-5mm}
  \caption{The architecture of the SELD network, which is a Conv-Conformer network. Dashed-yellow is the SED task. Dashed-blue is the DoA estimation task. The green boxes indicate soft connections between SED and DoA estimation. The convolution blocks can be extended to dense blocks. The network can be adapted to either log-mel spectrograms and IV input features, or SALSA input features. }
  \label{fig: single model}
\vspace*{-4mm}
\end{figure}

EINV2, which combines the track-wise output format, PIT, and soft PS, is utilized for our system. We extend EINV2 to three tracks to address up to three overlapped sound events. We then utilize Conformer to replace the multi-head self-attention (MHSA) blocks in EINV2. Conformer consists of two feed-forward layers with residual connections sandwiching the MHSA and convolution modules, where MHSA and convolution modules can capture global and local patterns, respectively. To increase the diversity of model ensembles, we replace convolution blocks with dense blocks. In dense blocks, each layer is connected with all preceding layers to obtain additional inputs, and delivers its own feature maps to all subsequent layers. This feed-forward structure can strengthen forward propagation of features and back propagation of gradients. Our proposed network is shown in Fig. \ref{fig: single model}. The detail of the network architecture can be found in the released code.

\subsection{Data Augmentation}
In practical applications, training set cannot cover all actual instances from different spatial and sound environments, and mismatches between the training set and test set are common. To improve the generalization of the model, we propose a novel data-augmentation method.

Our proposed data-augmentation is characterized by utilizing several augmentation operations\cite{hendrycks2019augmix, chen2020simple}. We randomly sample $k$ augmentation chains, where $k=3$ is used by default. Each augmentation chain is constructed by composing from some randomly selected augmentation operations. Augmentation operations that we used include Mixup\cite{zhang2018mixup}, SpecAugment\cite{specaug}, Cutout, and rotation of FOA signals\cite{mazzon2019first}.

Mixup trains a neural network on convex combinations of pairs of feature vectors and their labels. We use Mixup on both raw waveforms and features to improve the generalization for detecting overlapping sound events. While random Cutout produces several rectangular masks on features, SpecAugment produces time and frequency stripes to mask on features. We also use a spatial augmentation method, which is rotation of FOA signals. It rotates FOA format signals and enriches DoA labels without losing physical relationships between steering vectors and observer. We use x, y, and z axis as the rotation axis, respectively, which leads to 48 types of channel rotation.

\section{Post Processing}
\subsection{Track-wise Output Format}
The trackwise output format was introduced in previous works\cite{cao2020event,cao2021}. It can be defined as
\begin{equation}
\begin{split}
&\boldsymbol{Y}_{\text {Trackwise}}=\\ 
&\left\{\left(y_{\mathrm{SED}}, y_{\mathrm{DoA}}\right) \mid y_{\mathrm{SED}} \in \mathbb{O}_{\mathbf{S}}^{M \times K}, y_{\mathrm{DoA}} \in \mathbb{R}^{M \times 3}\right\}
\end{split}
\end{equation}
where $M$ is the number of tracks, $K$ is the number of  sound-event types, $\mathbb{O}_{\mathbf{S}}^{M \times K}$ is one hot encoding of $K$ classes, $\mathbf{S}$ is the set of sound events, and the number of dimensions of Cartesian coordinates is 3.

The number of tracks needs to be pre-determined according to the maximum number of overlapping events. Each track can only detect a sound event and a corresponding location. While a model with track-wise output format is trained, sound events are not always predicted in a fixed track. It may result in a problem that sound events predicted in a track may not be aligned to its ground truth. This may be due to the track permutation problem. Permutation-invariant training (PIT) can be utilized for the problem. The PIT loss is defined as

\begin{equation}
\begin{split}
&\mathcal{L}_{P I T}(t)=\\
&\min _{\alpha \in \mathbf{P}(t)} \sum_{M}\left\{\lambda\cdot\ell_{\mathrm{SED}}(t,\alpha )+(1-\lambda)\cdot \ell_{\mathrm{DoA}}(t,\alpha )\right\}
\end{split}
\end{equation}
where $\alpha\in \mathbf{P(t)}$ indicates one of the possible permutations and $\lambda$ is a weight between SED loss and DoA loss. $\ell_\mathrm{SED}$ is binary cross entropy loss for the SED task, and $\ell_{\mathrm{DoA}}$ is mean square error for the DoA task. The lowest loss will be chosen by finding a possible permutation, and the back-propagation is then performed.

\subsection{Track-wise Ensemble Model}

\begin{figure}[tb]
  \centering
  \scalebox{0.8}{\centerline{\includegraphics[width=\columnwidth]{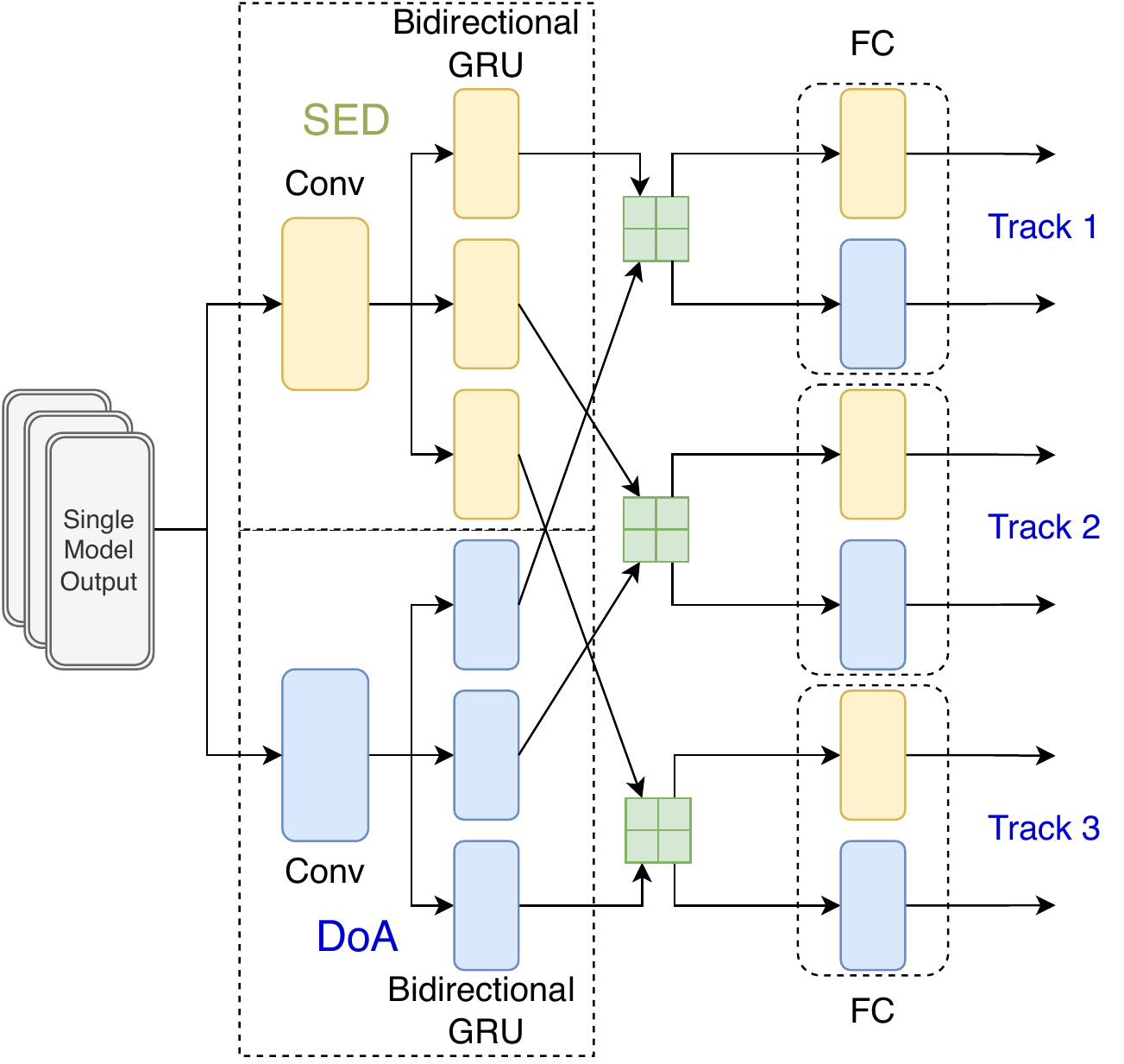}}}
  \vspace*{-5mm}
  \caption{The architecture of the proposed track-wise ensemble model}
  \label{fig: ensemble model}
\vspace*{-4mm}
\end{figure}

For the track-wise output format, the prediction of a sound event may be allocated to a track randomly. Methods like TTA, average or weighted ensemble\cite{nguyen2021dcase,shimada2021ensemble} cannot align predictions from different tracks, which makes these methods inapplicable to the track-wise output format. We propose a novel post-processing method named track-wise ensemble model. The track-wise ensemble model is a trainable model shown in Fig. \ref{fig: ensemble model}. The inputs to the ensemble model are the outputs from different single models. The ensemble model then predicts results in a manner of the track-wise output format. The model structure is a simplified version of EINV2 with SED and DoA estimation branches. Each branch is consisted of a convolutional-recurrent neural network (CRNN) with soft PS connecting two branches. CRNN consists of 2D convolution layers with 128 output channels and kernel size of $3\times3$, and bidirectional GRU with hidden size of 64.

After training $N$ inhomogeneous models, we get $N$ predictions $[ \mathbf{y}_\text{SED}^1,  \mathbf{y}_\text{SED}^2, \dots, \mathbf{y}_\text{SED}^N  ]$ and $[ \mathbf{y}_\text{DoA}^1,  \mathbf{y}_\text{DoA}^2, \dots, \mathbf{y}_\text{DoA}^N  ]$. We concatenate and flatten predictions of SED and DoA, which are then sent as the inputs to the ensemble model. The model also outputs track-wise format predictions.

\begin{table*}[!hbt]
\centering
\caption{The performance of our proposed model on the validation set}
\label{tab:perf_seld}
\begin{tabular}{cc|c|cc|cc}
\hline
\multirow{2}{*}{System} & \multirow{2}{*}{Models} & \multirow{2}{*}{Features} & \multicolumn{2}{c|}{$\text{F score}_{\leq\text{1m}}$} & \multicolumn{2}{c}{$\text{F score}_{\leq\text{2m}}$} \\ \cline{4-7} 
 &  &  & Single FOA     & Double FOAs & Single FOA & Double FOAs  \\ \hline
\#1                     & ConvBlock-Conformer     & log-mel + IV              & 0.667       & 0.677        & 0.695       & 0.700       \\
\#2                     & DenseBlock-Conformer    & log-mel + IV              & 0.653       & 0.668        & 0.674       & 0.689       \\
\#3                     & ConvBlock-Conformer     & SALSA                     & 0.651       & 0.661        & 0.681       & 0.685       \\ \hline
\end{tabular}
\end{table*}

\section{Experiments}
\label{sec:typestyle}
\subsection{Dataset}
We verify our proposed method using the dataset provided by the L3DAS22 Task 2\cite{l3das22}. The dataset is split into 3 subsets, which consist of 
600, 150, and 150 30-second-long audio recordings for the train, validation and test splits, respectively. There are 14 types of sound events that are selected from the FSD50K dataset. The maximum overlapping sound events are three. The room impulse response (RIR) is sampled in an office room with the dimension to be around 6 m (length) by 5 m (width) by 3 m (height). FOA microphone arrays are placed in the center of the room. The position of the FOA microphone arrays is set to be the origin of the coordinates.

\subsection{Hyper-parameters and Evaluation Metrics}

The sampling frequency of the dataset is 32 kHz. We used a 1024-point Hanning window with a hop size of 400 and 128 mel bins for log-mel spectrograms and IV features, and a 512-point Hanning window with a hop size of 400 for SALSA features. Audio clips are segmented to have a fix length of 5 seconds with no overlap for training. AdamW optimizer is used. The learning rate is set to 0.0003 for the first 90 epochs and is adjusted to 0.00003 for the following 10 epochs. The threshold for SED is set to 0.5 to binarize predictions. For two FOA microphone arrays, we extract 4 channels of log-mel spectrograms and 3 channels of IV features, and 4 channels of log-linear spectrograms and 3 channels of normalized principal eigenvectors for SALSA, respectively. Then we concatenate these features to make the input channels to be 14.

The evaluation metric uses a joint metric of localization and detection: F-score based on the location-sensitive detection. It counts true positives predicted when the label of a sound event is correct and its location is within a Cartesian distance threshold from its reference location\cite{mesaros2019joint}. 

\subsection{Experimental Results}

We trained the proposed model with different configurations using the training set of the L3DAS22 Task 2 dataset and evaluated the performance on the validation set. The configurations and results are shown in Table \ref{tab:perf_seld}. Our system is developed based on the spatial error threshold to be 1 m and is evaluated at the threshold to be both 1 m and 2 m. We also show the performance of our system with one FOA array and two FOA arrays. The performance of two FOA arrays is slightly better.

Table \ref{tab:ensemble} shows the performance of proposed data augmentation and ensemble model. We evaluate the performance of proposed data augmentation using the configuration of System \#1 in Table \ref{tab:perf_seld}. Several data-augmentation operations, which are directly linked in series, has a significant performance improvement compared to the method without any augmentations, but the method of augmentation chains performs much better. The average ensemble model is not stable and performs much worse than any of the single models that are shown in Table \ref{tab:perf_seld}. This is due to the track permutation problem. On the other hand, the track-wise ensemble model, which uses PIT to solve the track permutation problem among different models, can improve the performance compared with any single model.

\begin{table}[]
\caption{The performance of proposed data augmentation and ensemble model on the validation set. }
\label{tab:ensemble}
\begin{adjustbox}{width=\columnwidth,center}
\begin{tabular}{l|cc}
\hline
\multicolumn{1}{c|}{Methods}             & $\text{F score}_{\leq\text{1m}}$ & $\text{F score}_{\leq\text{2m}}$ \\ \hline
System \#1 without dataAug     & 0.515   & 0.566 \\
System \#1 with dataAug in series & 0.616   & 0.659 \\
System \#1 with dataAug in chains & \textbf{0.677}   & \textbf{0.700} \\ \hline
Average Ensemble    & 0.499   & 0.607   \\
Track-wise Ensemble & \textbf{0.685}   & \textbf{0.715}   \\ \hline
\end{tabular}
\end{adjustbox}
\end{table}

The final submitted system is trained on both L3DAS21 and L3DAS22 datasets. The evaluation results on the blind test set are shown in Table \ref{tab:results}. It can be seen that the performance of the proposed method outperforms the baseline method significantly on the blind test set.

\begin{table}[]
\caption{Evaluation results for submitted systems on the blind test set. The official spatial threshold is fixed to 2 m.}
\label{tab:results}
\begin{adjustbox}{width=\columnwidth,center}
\begin{tabular}{c|ccc}
\hline
\multirow{1}{*}{System}    & Precision & Recall & F score \\
\hline
Baseline                  & 0.423     & 0.289  & 0.343   \\
Track-wise ensemble model & \textbf{0.706}     & \textbf{0.691}  & \textbf{0.699}   \\ \hline
\end{tabular}
\end{adjustbox}
\end{table}


\section{Conclusion}
\label{sec:majhead}

We have presented a track-wise ensemble event independent network with a novel data augmentation approach for 3D polyphonic sound event localization and detection. The proposed data augmentation method contains several augmentation chains. Each augmentation chain contains several randomly sampled augmentation operations. In addition, the proposed ensemble model is based on single models that is extended from Event-Independent Network V2. We use log-mel spectrograms, intensity vectors, as well as Spatial Cues-Augmented Log-Spectrogram as input features to these single models. The performance of these single models is also improved by using conformer blocks and dense blocks. We adopt a trainable ensemble model with the track-wise output format to tackle with the track permutation problem for different models. Experimental results show that the proposed method can achieve location-dependent F-score of 0.699, which is the top ranking in the Task 2 of the L3DAS22 challenge. The proposed track-wise ensemble method can solve the track permutation problem well and outperforms the average ensemble method by a large margin.

\section{Acknowledgement}
\label{sec:print}

This work was supported in part by Frontier Exploration project independently deployed by Institute of Acoustics, Chinese Academy of Sciences (No. QYTS202009), National Natural Science Foundation of China (Grant No. 11804365), and EPSRC Grant EP/T019751/1 ``AI for Sound''.

\bibliographystyle{IEEEbib}
\bibliography{strings,refs}

\begin{thebibliography}{10}

\bibitem{Adavanne2018_JSTSP}
S.~Adavanne, A.~Politis, J.~Nikunen, and T.~Virtanen,
\newblock ``{Sound event localization and detection of overlapping sources
  using convolutional recurrent neural networks},''
\newblock {\em IEEE J Sel Top Signal Process}, vol. 13, pp. 34--48, 2018.

\bibitem{Adavanne2019_DCASE}
S.~Adavanne, A.~Politis, and T.~Virtanen,
\newblock ``A multi-room reverberant dataset for sound event localization and
  detection,''
\newblock in {\em Proc. DCASE 2019 Workshop}, 2019, pp. 10--14.

\bibitem{Cao2019}
Y.~Cao, Q.~Kong, T.~Iqbal, F.~An, W.~Wang, and M.~D. Plumbley,
\newblock ``Polyphonic sound event detection and localization using a two-stage
  strategy,''
\newblock in {\em Proc. DCASE 2019 Workshop}, 2019, pp. 30--34.

\bibitem{cao2020event}
Y.~Cao, T.~Iqbal, Q.~Kong, Y.~Zhong, W.~Wang, and M.~D. Plumbley,
\newblock ``Event-independent network for polyphonic sound event localization
  and detection,''
\newblock in {\em Proc. DCASE 2020 Workshop}, 2020, pp. 11--15.

\bibitem{wang2021four}
Q.~Wang, J.~Du, H.~Wu, J.~Pan, F.~Ma, and C.~Lee,
\newblock ``A four-stage data augmentation approach to {ResNet-Conformer} based
  acoustic modeling for sound event localization and detection,''
\newblock {\em arXiv preprint arXiv:2101.02919}, 2021.

\bibitem{perotin2019crnn}
L.~Perotin, R.~Serizel, E.~Vincent, and A.~Guerin,
\newblock ``{CRNN}-based multiple {DoA} estimation using acoustic intensity
  features for ambisonics recordings,''
\newblock {\em IEEE J Sel Top Signal Process}, pp. 22--33, 2019.

\bibitem{celsi2020quaternion}
M.~Ricciardi~Celsi, S.~Scardapane, and D.~Comminiello,
\newblock ``Quaternion neural networks for {3D} sound source localization in
  reverberant environments,''
\newblock in {\em Proc. IEEE MLSP 2020}, 2020, pp. 1--6.

\bibitem{nguyen2021dcase}
T.~Nguyen, K.~Watcharasupat, N.~K. Nguyen, D.~L. Jones, and W.~S. Gan,
\newblock ``{DCASE} 2021 {Task} 3: Spectrotemporally-aligned features for
  polyphonic sound event localization and detection,''
\newblock {\em arXiv preprint arXiv:2106.15190}, 2021.

\bibitem{nguyen2021salsa}
T.~Nguyen, K.~Watcharasupat, N.~K. Nguyen, D.~L. Jones, and W.~S. Gan,
\newblock ``\text{SALSA}: Spatial cue-augmented log-spectrogram features for
  polyphonic sound event localization and detection,''
\newblock {\em arXiv preprint arXiv:2110.00275}, 2021.

\bibitem{cao2021}
Y.~Cao, T.~Iqbal, Q.~Kong, F.~An, W.~Wang, and M.~D. Plumbley,
\newblock ``An improved event-independent network for polyphonic sound event
  localization and detection,''
\newblock in {\em Proc. IEEE ICASSP 2021}, 2021, pp. 885--889.

\bibitem{huang2017densely}
G.~Huang, Z.~Liu, L.~Maaten, and K.~Q. Weinberger,
\newblock ``Densely connected convolutional networks,''
\newblock in {\em Proc. IEEE CVPR 2017}, 2017, pp. 4700--4708.

\bibitem{gulati2020conformer}
A.~Gulati, J.~Qin, C.~Chiu, N.~Parmar, Y.~Zhang, J.~Yu, W.~Han, S.~Wang,
  Z.~Zhang, Y.~Wu, and R.~Pang,
\newblock ``Conformer: Convolution-augmented transformer for speech
  recognition,''
\newblock in {\em Proc. Interspeech 2020}, 2020, pp. 5036 -- 5040.

\bibitem{mazzon2019first}
L.~Mazzon, Y.~Koizumi, M.~Yasuda, and N.~Harada,
\newblock ``First order ambisonics domain spatial augmentation for {DNN}-based
  direction of arrival estimation,''
\newblock in {\em Proc. DCASE 2019 Workshop}, 2019, pp. 154--158.

\bibitem{shimada2021ensemble}
K.~Shimada, N.~Takahashi, Y.~Koyama, S.~Takahashi, E.~Tsunoo, M.~Takahashi, and
  Y.~Mitsufuji,
\newblock ``Ensemble of {ACCDOA}- and {EINV2}-based systems with {D3Nets} and
  impulse response simulation for sound event localization and detection,''
\newblock in {\em arXiv preprint arXiv:2106.10806}, 2021.

\bibitem{wolpert1992stacked}
D.~H. Wolpert,
\newblock ``Stacked generalization,''
\newblock {\em Neural Networks}, vol. 5, pp. 241--259, 1992.

\bibitem{l3das22}
E.~Guizzo, C.~Marinoni, M.~Pennese, X.~Ren, X.~Zheng, C.~Zhang, B.~Masiero,
  A.~Uncini, and D.~Comminiello,
\newblock ``{L3DAS22 Challenge}: Learning {3D} audio sources in a real office
  environment,''
\newblock in {\em Proc. IEEE ICASSP 2022}, 2022.

\bibitem{hendrycks2019augmix}
D.~Hendrycks, N.~Mu, E.~D. Cubuk, B.~Zoph, J.~Gilmer, and B.~Lakshminarayanan,
\newblock ``{AugMix}: A simple data processing method to improve robustness and
  uncertainty,''
\newblock in {\em Proc. ICLR 2020}, 2020.

\bibitem{chen2020simple}
T.~Chen, S.~Kornblith, M.~Norouzi, and G.~Hinton,
\newblock ``A simple framework for contrastive learning of visual
  representations,''
\newblock in {\em Proc. ICML 2020}, 2020, pp. 1597--1607.

\bibitem{zhang2018mixup}
H.~Zhang, M.~Cisse, Y.~N. Dauphin, and D.~Lopez-Paz,
\newblock ``mixup: Beyond empirical risk minimization,''
\newblock in {\em Proc. ICLR 2018}, 2018.

\bibitem{specaug}
D.~S. Park, W.~Chan, Y.~Zhang, C.~Chiu, B.~Zoph, E.~D. Cubuk, and Q.~V. Le,
\newblock ``{SpecAugment}: A simple data augmentation method for automatic
  speech recognition,''
\newblock in {\em Proc. Interspeech 2019}, 2019, pp. 2613 -- 2617.

\bibitem{mesaros2019joint}
A.~Mesaros, S.~Adavanne, A.~Politis, T.~Heittola, and T.~Virtanen,
\newblock ``Joint measurement of localization and detection of sound events,''
\newblock in {\em Proc. IEEE WASPAA 2019}, 2019, pp. 333--337.

\end{thebibliography}

\end{document}